\documentclass[12pt]{article}

\usepackage{amssymb}
\usepackage{amsmath}
\usepackage{latexsym}
\catcode`@=11
\renewcommand{\section}{\@startsection{section}{1}{0pt}{\medskipamount}
{\medskipamount}{\large\bf}}
\numberwithin{equation}{section}
\catcode`@=12

\def\beq{\begin{eqnarray}}    
\def\eeq{\end{eqnarray}}      

\def\ln{\,\mbox{ln}\,}                  
\def\sTr{\,\mbox{sTr}\,}                
\def\sDet{\,\mbox{sDet}\,}              

\def\={\ =\ }

\begin{document}



\begin{center}

{\large\bf A SYSTEMATIC STUDY OF FINITE BRST-BFV TRANSFORMATIONS IN  GENERALIZED
HAMILTONIAN FORMALISM}

\vspace{18mm}

{\Large Igor A. Batalin$^{(a)}\footnote{E-mail:
batalin@lpi.ru}$\;,
Peter M. Lavrov$^{(b, c)}\footnote{E-mail:
lavrov@tspu.edu.ru}$\;,
Igor V. Tyutin$^{(a)}\footnote{E-mail: tyutin@lpi.ru}$
}

\vspace{8mm}


\noindent ${{}^{(a)}}$
{\em P.N. Lebedev Physical Institute,\\
Leninsky Prospect \ 53, 119 991 Moscow, Russia}

\noindent  ${{}^{(b)}}
${\em
Tomsk State Pedagogical University,\\
Kievskaya St.\ 60, 634061 Tomsk, Russia}

\noindent  ${{}^{(c)}}
${\em
National Research Tomsk State  University,\\
Lenin Av.\ 36, 634050 Tomsk, Russia}

\vspace{20mm}

\begin{abstract}
\noindent
We study systematically finite BRST-BFV transformations in the generalized
Hamiltonian formalism.
We present explicitly their Jacobians and the form of a solution to the
compensation equation determining the functional field dependence of finite
Fermionic  parameters, necessary to
generate arbitrary  finite change of gauge-fixing functions in the path integral.
\end{abstract}

\end{center}

\vfill




\section{INTRODUCTION}

It is well-known that BRST-BFV symmetry \cite{brs1,brs2,t,FV,BVhf}
is a powerful tool to study general properties of gauge field
systems \cite{FF,BV,VLT,LT,BF,H,HT}. Parameters of that symmetry are
constant Fermions, although they are allowed to be functionals of
fields. Usually, the symmetry is introduced infinitesimally, which
means that its Fermionic parameters are considered formally as
infinitely-small quantities. Usual strategy is to show that the
Jacobian of BRST-BFV transformation does generate arbitrary
variation of gauge-fixing functions in the path integral. This can
be done by choosing necessary functional dependence of BRST-BFV
parameters on fields.

For Yang-Mills theories in Lagrangian formalism based on the
Faddeev-Popov method \cite{FP} a study of finite field-dependent
BRST transformations was initiated in Ref. \cite{JM} where a
differential equation for the Jacobian of such change of variables
in vacuum functional has been proposed. But a solution to this
equation has not been found in Ref. \cite{JM} and numerous further
studies of this problem. Recently \cite{LL}, it was proved that the
problem of finding an explicit form of the Jacobian is pure
algebraic and can be solved in terms of the BRST variation of
field-dependent parameter. Any finite field dependent BRST
transformation  of variables in the generating functional of Green
functions is related to modification of gauge fixing functional
\cite{LL,LL1}.

By following the ideas of paper \cite{JM}, recently in Ref.
\cite{RM} an attempt was made to consider finite field-dependent
BRST transformations in Hamiltonian formalism within BFV
quantization method of dynamical systems with constraints
\cite{FV,BVhf,FF}. The same as in Ref. \cite{JM}, the main result
was formulated here as a differential equation for the Jacobian of
these transformations. Again, an explicit solution to the
differential equation was not found in this approach.

In the present article, we develop systematically the concept of finite BRST-BFV
transformations,
which means actually that we consider formally BRST-BFV parameters as finite
Fermionic quantities. Historically, there were several authors
(see \cite{FF,BV,VLT,LT,BF,H,HT} and references herein)
who worked  sporadically with
finite BRST-BFV
transformations. But the final results were formulated  infinitesimally
 even in these special cases. In the present work, we give a unique consistent approach.
Thereby, our new strategy is to show that the Jacobian of these
finite transformations does generate an arbitrary finite change of
gauge-fixing functions in the path integral. In order to do this, we
formulate the compensation equation determining  the necessary
functional field dependence for finite Fermionic parameters. Then, we
present the explicit solution to that compensation equation. We find
functional formulation of BRST-BFV transformations and derive the
Ward identities as well as a relation connecting generating
functionals of Green functions written in two different gauges.
\\

\section{FINITE BRST-BFV TRANSFORMATIONS AND THEIR \\
JACOBIANS}

Let
\beq
z^{i}=( q ; p ),\;\; \varepsilon(z^i)=\varepsilon_i,
\eeq
be a complete set of canonical variables specific to the extended
phase space of generalized Hamiltonian formalism. The partition
function reads
\beq
\label{E2}
Z_\psi = \int Dz \exp\left[\left(\frac{i}{\hbar}\right)W_\psi\right],
\eeq
where the action $W_\psi$ is defined as
\beq
 \label{E3}
&& W_\psi=\int\left[\left(\frac{1}{2}\right)z^i(t)\omega_{ik}\dot{z}^k(t)-H(t)\right]dt,
\quad
H(t)=\mathcal{H}(t)+\{\Omega,\psi\}_t,  \\
 \label{E4}
&& \{\Omega,\Omega\} = 0, \quad \{\Omega,\mathcal{H}\}=0,\\
\label{E5}
 &&                     
\mathcal{H} = H_{0} + ...     , \quad \Omega = c T + ....      ,                 
\eeq
Here in (\ref{E3}), $z^i(t)$ are functions of
time (trajectories), $\dot{z}^k(t)=dz^k(t)/dt$, $H(t),\mathcal{H}(t)$,
$\Omega(t),\psi(t)$ are local functions  of time:
$H(t)=\left.H(z)\right|_{z\rightarrow z(t)}$ and so on, $\{,\}_t$ means the
Poisson superbracket for fixed time $t$:
$\{\Omega,\psi\}_t=\left.\{\Omega(z),\psi(z)\}\right|_{z\rightarrow z(t)}$
and so on, $\{ z^{i}, z^{k} \} = \omega^{ik} = \hbox{const} = -
\omega^{ki}(-1)^{\varepsilon_i\varepsilon_k}$ is an invertible even
supermatrix; $\omega_{ik}$ ($\omega_{ik}=
(-1)^{(\varepsilon_i+1)(\varepsilon_k+1)}\omega_{ki}$) stands for an inverse
to $\omega^{ik}$, $\mathcal{H}$ is a
Boson with ghost number zero, while $\Omega$ and $\psi$  is a Fermion with
ghost number $+1$ and $-1$, respectively, they are called BRST-BFV generator
and gauge Fermion.

It follows from (\ref{E4}) that
\beq
\label{E6}
\{\Omega,H\} = 0.                                                          
\eeq
We define finite BRST-BFV transformations of phase (canonical) variables
$z^k\rightarrow {\bar z}^k$ in the form
\beq
\label{E7}
{\bar z}^k= z^k + \{z^k,\Omega\}\mu,                                             
\eeq
and then finite BRST-BFV transformations of the trajectories
$z^k(t)\rightarrow{\bar z}^k(t)$,
\beq
\label{E7a}
{\bar z}^k(t)=\left.{\bar z}^k\right|_{z\rightarrow z(t)}= z^k(t) +
\{z^k,\Omega\}_t\mu,                                             
\eeq
where $\mu$ is a finite  Fermionic parameter.
In general, $\mu=\mu[z]$ is a functional of the whole trajectory
${z^k(t),\; -\infty < t < +\infty }$. However, $\mu$ itself is independent of
current time $t$ and phase variables $z^k$,
\beq
\label{E8}
d_{t}\mu = 0, \; \partial_{k} \mu =0,                                      
\eeq
where we have denoted
\beq
\label{E9}
d_{t} = d / dt , \; \partial_{k}=\partial/\partial z^{k}.                         
\eeq
Thus, only a functional derivative $\delta/\delta z(t)$ is capable
to differentiate $\mu[z]$ nontrivially.

Now, let us consider the functional Jacobian,
\beq
&&J=\sDet\left\{{\bar z}^i(t)\frac{\overleftarrow{\delta}}{\delta z^j(t')}\right\}=
\exp\left\{\sTr\ln\left[{\bar z}^i(t)
\frac{\overleftarrow{\delta}}{\delta z^j(t')}\right]\right\}= \\
\nonumber
\label{E10}  
&&=\exp\left\{\sTr\ln\left[\delta^i_j\delta(t-t^\prime)+
\{z^i,\Omega\}_t\left(\mu\frac{\overleftarrow{\delta}}{\delta z^{j}(t')}\right)+
(\{z^i,\Omega\}\overleftarrow{\partial_j})_t(-1)^{\varepsilon_j}\mu\delta(t-t^\prime)\right]
\right\}.
\eeq
We factorize the Jacobian (\ref{E10}) in the form
\beq
\label{E11}
J = J_{1} J_{2},                                                                
\eeq
where
\beq
\label{E12}
&&J_1=\exp\big\{\sTr\ln(G^{-1})^i_k(t,t';\lambda=1)\big\},\\
&&\;\;(G^{-1})^i_k(t,t';\lambda)=\delta^i_k\delta(t-t^\prime)+
\lambda\{z^i,\Omega\}_t
\left(\mu\frac{\overleftarrow{\delta}}{\delta z^{k}(t')}\right)\;,
\\
\label{E13}                                                                  
&&J_2= \exp\big\{\sTr\ln[\delta^i_k\delta(t-t')+(\{z^i,\Omega\}
\overleftarrow{\partial_j})_t
G^j_k(t,t';\lambda=1)(-1)^{\varepsilon_k}\mu]\big\}\;.
\eeq
It follows from an equation for matrix $G^i_k(t,t';\lambda)$,
\beq \label{E13a}
G^i_k(t,t';\lambda)+\lambda\{z^i,\Omega\}_t
\int dt''\left(\mu\frac{\overleftarrow{\delta}}{\delta z^j(t'')}\right)G^j_k(t'',t';\lambda)=
\delta^i_j\delta(t-t^\prime),
\eeq
that
\beq
\label{E14}
&&                                                                                  
G^i_k(t,t';\lambda)=\delta^i_k\delta(t-t')-\lambda\{z^i,\Omega\}_tA_k(t'),  \\
\nonumber                                        
&&A_k(t')=\int dt''\mu\frac{\overleftarrow{\delta}}{\delta z^j(t'')}
G^j_k(t'',t';\lambda)=\big(1+\lambda\kappa\big)^{-1}
\left(\mu\frac{\overleftarrow{\delta}}{\delta z^k(t')}\right), \\
\label{E15}
&&
\kappa=\int dt''\mu\frac{\overleftarrow{\delta}}{\delta z^i(t'')}\{z^i,\Omega\}_{t''}.                                      
\eeq
It is a characteristic feature of the factor (\ref{E12}) that the operator
therein is nontrivial only for $\mu$ depending actually on fields. On the
other hand, in the factor (\ref{E13}), the corresponding operator has a nontrivial
part  proportional to undifferentiated $\mu$, so that the $\mu$-power series
expansion terminates at the quadratic order, as $\mu$
itself is nilpotent due to its Fermionic nature. Let us consider the factors
(\ref{E12}), (\ref{E13}) in more detail.

For the $J_{1}$ factor, we have
\beq
\nonumber
&&\ln J_{1} = \int_0^1d\lambda\int dtdt'G^i_j(t, t';\lambda)\{z^j,\Omega\}_{t'}
\left(\mu\frac{\overleftarrow{\delta}}{\delta z^i(t)}\right)(-1)^{\varepsilon_i} =\\
\label{E16}
&&
=-\int_0^1d\lambda\int dt'A_j(t')\{z^j,\Omega\}_{t'}=
-\int_0^1d\lambda\big(1+\lambda\kappa\big)^{-1}\kappa=-\ln(1+\kappa).                                      
\eeq
Then, for the factor $J_{2}$, we have
\beq
\nonumber
&&\ln J_{2}=\sTr[(\{z^i,\Omega\}\overleftarrow{\partial}_k)_t
G^k_j(t,t';\lambda=1)(-1)^{\varepsilon_j}]\mu = \\
\label{E17}
&&=\sTr\left\{[(\{z^i,\Omega\}\overleftarrow{\partial}_j)_t\delta(t-t')-
(\{z^i,\Omega\}\overleftarrow{\partial}_k\{z^k,\Omega\})_t
A_j(t')](-1)^{\varepsilon_j}\right\}\mu.
\eeq

Here in the square brackets in (\ref{E17}), the first term is zero due to
the Liouville's theorem,
\beq
\label{E18}
\{z^i,\Omega\}\overleftarrow{\partial}_i =
\omega^{ik}\partial_k\Omega\overleftarrow{\partial}_i =
\omega^{ik}\partial_k\partial_i\Omega=0.                                       
\eeq
The second term in the square bracket in (\ref{E17}) is zero due to the Jacobi identity
and the first in (\ref{E4}),
\beq
\label{E19}
\{\{z^i,\Omega\},\Omega\}=\left\{z^i,\left(\frac{1}{2}\right)\{\Omega,\Omega\}\right\}=0.       
\eeq

Thus, we arrive at
\beq
\label{E20}
J = J_1 = \exp\big\{-\ln(1+\kappa)\big\}=(1+\kappa)^{-1}
\eeq
for the functional Jacobian  (\ref{E10}) of the finite
BRST-BFV transformation (\ref{E7}).
\\

\section{COMPENSATION EQUATION AND ITS \\
EXPLICIT SOLUTION   }

Now, we would like to use the Jacobian (\ref{E20}) to generate arbitrary finite
change $\delta \psi$
of the gauge Fermion $\psi$ in the action (\ref{E3}),
\beq
\label{E21}
\psi \rightarrow \psi_1=\psi + \delta \psi.                                                 
\eeq

Let us make the transformation (\ref{E7a}) of the trajectories in the path integral
(\ref{E2}) for partition function.

First of all, the action (\ref{E3}) in the new variables (\ref{E7a}) reads
\beq
\label{E22}
{\bar W}_{\psi} = \int \left[\left(\frac{1}{2}\right){\bar z}^{i}(t)\omega_{ik}d
{\bar z}^{k}(t)/dt-
{\bar H}(t)\right]dt= W_{\psi}          
\eeq
where we have used
\beq
\label{E23} 
\int[{\bar z}^i(t)\omega_{ik}d{\bar z}^k(t)/dt]dt=\int[z^i(t)\omega_{ik}dz^k(t)/dt]dt+
\left.(z^k\partial_k\Omega-2\Omega)_t\mu\right|_{-\infty}^{+\infty},
\eeq
and ${\bar H}=H({\bar z})=H(z)$ (with eq. (\ref{E6}) taken into account).
Now, we have for the partition  function in the new variables,
\beq
&& \nonumber
Z_\psi=\int D{\bar z}\exp\left[\left(\frac{i}{\hbar}\right){\bar W}_{\psi}\right]=
\int DzJ\exp\left[\left(\frac{i}{\hbar}\right)W_\psi\right]= \\
\label{E25}
&&=\int Dz\exp\left\{\left(\frac{i}{\hbar}\right)\left[W_{\psi_1}-
\left(-\int dt\{\Omega,\delta\psi\}_t+
\left(\frac{\hbar}{i}\right)\ln(1+\kappa)\right)\right]\right\}.
\eeq
Let us require the following condition to hold
\beq
\label{E26a}
J=\exp\left[-\left(\frac{i}{\hbar}\right)\int dt\{\Omega,\delta\psi\}_t\right].                        
\eeq
It follows then
\beq
\label{E27}
Z_{\psi_1}=Z_\psi,                                      
\eeq
for arbitrary finite $\delta\psi$. We call the condition (\ref{E26a})
a "compensation equation".
Due to the formula (\ref{E20}), it follows that (\ref{E26a}) is rewritten  as
\beq
\label{E28}
\int dt \mu\frac{\overleftarrow{\delta}}{\delta z^i(t)}\{z^i,\Omega\}_t
=\exp\left[\left(\frac{i}{\hbar}\right)\int dt\{\Omega,\delta\psi\}_t\right] - 1.                            
\eeq
That is a functional equation to determine $\mu[z]$. Introducing a functional $x$,
\beq \label{E28a} x=\left(\frac{i}{\hbar}\right)\int
dt\{\Omega,\delta\psi\}_t=\left(\frac{i}{\hbar}\right)\delta\Psi\int dt
\frac{\overleftarrow{\delta}}{\delta z^i(t)}\{z^i,\Omega\}_t,\quad
\delta\Psi=\int dt\delta\psi(t),
\eeq
we can rewrite eq. (\ref{E28}) in the form
\beq \label{E28b}
\mu\int dt\frac{\overleftarrow{\delta}}{\delta z^i(t)}\{z^i,\Omega\}_t=
f(x)x=\left(\frac{i}{\hbar}\right)[f(x)\delta\Psi]\int dt
\frac{\overleftarrow{\delta}}{\delta z^i(t)}\{z^i,\Omega\}_t\;,                          
\eeq
where $f(x)=\big(\exp(x)-1\big)\,x^{-1}$.
There is an obvious explicit solution to that equation
\beq
\label{E29}
\mu[\delta\psi]=\mu[z;\delta\psi]=\left(\frac{i}{\hbar}\right)f(x)\delta\Psi. 
\eeq
Thus we have confirmed explicitly the compensation equation
(\ref{E28}) to hold.

In the first order in $\delta\psi$, explicit solution (\ref{E29}) takes the
 usual form
\beq
\label{E31}
 \mu[\delta\psi] =\left(\frac{i}{\hbar}\right)\delta\Psi+O\big((\delta\psi)^2\big).                                        
\eeq

\section{FUNCTIONAL BRST-BFV TRANSFORMATIONS FOR \\
TRAJECTORIES
}

It appears quite natural to make our considerations above more transparent
by introducing
a  concept of functional BRST-BFV transformations. Namely, let us define
 a functional operator (differential) $\overleftarrow{d}$
of the form
\beq
\label{E32}
\overleftarrow{d}=\int dt\frac{\overleftarrow{\delta}}{\delta z^{i}(t)}
\{ z^{i},\Omega\}_{t}\;\qquad \varepsilon(\overleftarrow{d})=1\;.    
\eeq
It follows from the second in (\ref{E4}) that the Fermionic operator
(\ref{E32}) is nilpotent,
\beq
\label{E33}
\overleftarrow{d}^{2} = \left(\frac{1}{2}\right) [ \overleftarrow{d}, \overleftarrow{d} ] =
0\;.                                        
\eeq
The transformation (\ref{E7}) can be rewritten in terms of the operator (\ref{E32}) as
\beq
\label{E34}
{\bar z}(t) = z(t)( 1 + \overleftarrow{d} \mu) .                                   
\eeq
Thus, the operator (\ref{E32}) is a functional BRST-BFV generator.
Now, let us define the transformed action,
\beq
\label{E35}
{\bar W}_{\psi}=W_\psi(1+\overleftarrow{d}\mu).                                       
\eeq
Then, we get exactly the formula (\ref{E22}),
\beq
\label{E36}
{\bar W}_{\psi}=W_\psi+\left(\frac{1}{2}\right)(z^k\partial_k\Omega-2\Omega)_t\mu
\big|^{+\infty}_{-\infty}=W_\psi. 
\eeq

As $\overleftarrow{d}$ is linear and $\mu$ is nilpotent, applying the
operator $1+\overleftarrow{d}\mu$ to arbitrary functional $F(z)$,
$\overline{F}(z)=F(z)(1+\overleftarrow{d}\mu)$, yields the result
$\overline{F}(z)=F(\overline{z})$.

Functional Jacobian (\ref{E20}) is rewritten  in terms of the generator
 (\ref{E32}) as
\beq
\label{E37}
J=[1+(\mu\overleftarrow{d})]^{-1}.                                              
\eeq
The $x$ in (\ref{E28a}) can be represented as
\beq
\label{E39}
x=\left(\frac{i}{\hbar}\right)\delta\Psi\overleftarrow{d},                               
\eeq
and the compensation equation (\ref{E28}) takes the form
\beq
\label{E38}
\mu\overleftarrow{d}=\exp\left[\left(\frac{i}{\hbar}\right)
(\delta\Psi\overleftarrow{d})\right]-1=
\left(\frac{i}{\hbar}\right)[f(x)\delta\Psi]\overleftarrow{d}.    
\eeq

Thus, we conclude that all the main objects in our considerations can be
expressed naturally in terms of a single quantity that is the functional
BRST-BFV generator (\ref{E32}).

Note that the introduced transformations (\ref{E7a}), (\ref{E34}) form a group.
Indeed, let us rewrite (\ref{E34}) in the form
\beq
\label{E1a}
\overline{z}=z\overleftarrow{T}(\mu), \qquad \overleftarrow{T}(\mu)=1+
\overleftarrow{d}\mu\;.   
\eeq
then the composition law of transformations (\ref{E1a}) reads
\beq
\label{E2a}
\overleftarrow{T}(\mu_1)\overleftarrow{T}(\mu_2)=\overleftarrow{T}(\mu_{12}),
\qquad  \mu_{12}=\mu_1+J_{\mu_1}^{-1}\mu_2,     
\eeq
where  $J_{\mu_1}$ is the Jacobian of the transformation (\ref{E1a})
with $\mu_1$ standing for $\mu$. Indeed,  due to the nilpotency (\ref{E33}), we have
\beq
\label{E3a}
\overleftarrow{T}(\mu_1)\overleftarrow{T}(\mu_2)=1+\overleftarrow{d}\mu_1+
\overleftarrow{d}\mu_2+\overleftarrow{d}\mu_1\overleftarrow{d}\mu_2=
1+\overleftarrow{d}\mu_1+\overleftarrow{d}\mu_2+
\overleftarrow{d}(\mu_1\overleftarrow{d})\mu_2. 
\eeq
By substituting here
\beq
\mu_1\overleftarrow{d}=J_{\mu_1}^{-1}-1,              
\eeq
we arrive at (\ref{E2a}). Moreover, it follows from (\ref{E2a}) the relation
\beq
[\overleftarrow{d}\mu_1,\overleftarrow{d}\mu_2] =
\overleftarrow{d}\mu_{[12]}\;, \qquad
\mu_{[12]}=\mu_{12}-\mu_{21}=-(\mu_1\mu_2)\overleftarrow{d}.
\eeq

\section{WARD IDENTITIES DEPENDENT OF BRST-BFV \\
PARAMETERS/FUNCTIONALS}

As we have defined finite BRST- BFV transformations, it appears quite natural
to apply them
immediately to deduce the corresponding modified version  of the Ward identity.
We will do
that just in terms of functional BRST- BFV generator introduced in Section 4.

As usual for that matter, let us proceed with the external-source
dependent generating functional,
\beq
\label{E40}
Z_{\psi}(\zeta)=\int Dz\exp\left[\left(\frac{i}{\hbar}\right)W_{\psi}(\zeta)\right],\qquad
W_{\psi}(\zeta)=W_{\psi}+\int dt\zeta_k(t)z^k(t),           
\eeq
where $\zeta_k(t)$ ($\varepsilon(\zeta_k)=\varepsilon_k$)
is an arbitrary external source.
Of course, in the presence of non-zero external source,
the path integral (\ref{E40})  is in general actually dependent of gauge
Fermion $\psi$. However, due to the equivalence theorem \cite{KT}, this dependence
has a special form so that physical quantities do not depend on gauge.
In its turn, the Ward identity  measures the deviation of the path
integral from being  gauge-independent.

Let us perform in (\ref{E40})
the change  $z^{i}\rightarrow \overline{z}^{i}$
of integration variables, where $\overline{z}$  is defined by (\ref{E34}) )
with arbitrary $\mu[z]$. Then, by using the gauge invariance (\ref{E22})
as well as the (\ref{E37}) for the Jacobian, we get what we call a
"modified Ward identity",
\beq
\label{E41}
\left<\left[1+
\left(\frac{i}{\hbar}\right)\int dt\zeta_k(t)(z^k(t)\overleftarrow{d})\mu\right]
[1+(\mu\overleftarrow{d})]^{-1}\right>_{\psi,\zeta} = 1,                     
\eeq
where we have denoted the source dependent mean value
\beq
\label{E42}
\langle ( ... )\rangle_{\psi,\zeta}=\big[Z_{\psi}(\zeta)\big]^{-1}
\int Dz(...)\exp\left[\left(\frac{i}{\hbar}\right)W_{\psi}(\zeta)\right],\qquad
\langle 1 \rangle_{\psi,\zeta}=1,                     
\eeq
related to the source dependent action in the second in (\ref{E40}).
By construction, in (\ref{E41}) both $\zeta_{i}(t)$
and $\mu[z]$ are arbitrary. The presence of arbitrary $\mu[z]$ in the
integrand in (\ref{E41}) reveals the implicit dependence of the
generating functional (\ref{E40}) on the gauge-fixing Fermion $\Psi$
for nonzero external source $\zeta_{i}$.

For a constant $\mu$, $\mu =\hbox{const}$, the latter does drop-out
completely, and  we get  from (\ref{E41})
\beq
\label{E43}
\left<\int dt\zeta_k(t)(z^k(t)\overleftarrow{d})\right>_{\psi,\zeta} = 0,        
\eeq
which is exactly the standard form of a Ward identity. By identifying the
$\mu$ in (\ref{E41}) with the solution of the compensation equation
(\ref{E38}), it follows according to our result in Sec. 3,
\beq
\label{E44}
Z_{\psi_1}(\zeta) = Z_\psi(\zeta)\left[1+\left<\left(\frac{i}{\hbar}\right)
\int dt\zeta_k(t)(z^k(t)\overleftarrow{d})\mu[-\delta\psi]
\right>_{\psi,\zeta}\;\right]\!.
\eeq

Formula (\ref{E44}) generalizes (\ref{E27}) to the  presence of the external
sources.

Finally, let us notice the following. If we introduce the so-called
"antifields" $z^*_i(t)$, whose statistics is opposite to that of $z^{i}(t)$,
by adding the term $\int dt z^*_k(t)(z^k(t)\overleftarrow{d})$ to the second in
(\ref{E40}), we get the new generating functional $Z_\psi(\zeta,z^*)$.
\beq
\label{E46}
 Z_\psi(\zeta,z^*)=\int Dz\exp\left[\left(\frac{i}{\hbar}\right)W_\psi(\zeta,z^*)\right]\;
 ,\qquad
 W_\psi(\zeta,z^*)=W_\psi(\zeta)+\int dtz^*_k(t)\big(z^k(t)\overleftarrow{d}\big).     
\eeq
Now, let us perform in (\ref{E46}) the change $z^i\rightarrow\overline{z}^i$ with
arbitrary $\mu[z]$, we get
\beq
\label{E47aa}
\left<\left[1+\left(\frac{i}{\hbar}\right)\int dt\zeta_k(t)(z^k(t)\overleftarrow{d})
\mu\right]
[1+(\mu \overleftarrow{d})]^{-1}\right>_{\psi,\zeta,z*}=1,           
\eeq
where
\beq
\label{E48}
\langle(...)\rangle_{\psi,\zeta,z^*}=\big[Z_\psi(\zeta,z^*)\big]^{-1}\int Dz(...)
\exp\left[\left(\frac{i}{\hbar}\right)
W_\psi(\zeta,z^*)\right]\;, \qquad  \langle 1 \rangle_{\psi,\zeta,z^*}= 1. 
\eeq
Due to the nilpotency of $\overleftarrow{d}$, the only difference between (\ref{E41}) and
(\ref{E47aa}) is the one between (\ref{E42}) and (\ref{E48}). For a
$\mu=\hbox{const}$, we get from (\ref{E47aa})
\beq
\label{E49}
\left<\int dt\zeta_k(t)(z^k(t)\overleftarrow{d})\right>_{\psi,\zeta,z^*} = 0.
\eeq

In terms of (\ref{E46}), the latter is rewritten in a variation-derivative form
\beq
\label{E50}
\int dt \zeta_k(t)\frac{\overrightarrow{\delta}}{\delta z^*_k(t)}
\ln Z_\psi(\zeta,z^*) = 0\;.
\eeq
Now, let $S(z,z^*)$ be a functional Legendre transform to
$(\hbar/i)\ln Z_\psi(\zeta,z^*)$  with respect to the external source
$\zeta_{i}$,
\beq
&&z^k = \left(\frac{\hbar}{i}\right)\ln Z_\psi(\zeta,z^*)
\frac{\overleftarrow{\delta}}{\delta\zeta_k}(-1)^{\varepsilon_k}
=\left(\frac{\hbar}{i}\right)\frac{\overrightarrow{\delta}}{\delta\zeta_k}
\ln Z_\psi(\zeta, z^*), \\     
&&S(z,z^*)=\left(\frac{\hbar}{i}\right)\ln Z_\psi(\zeta,z^*)-\int dt \zeta_k(t)z^k(t),  \\
\label{E47}          
&&S(z,z^*)\frac{\overleftarrow{\delta}}{\delta z^j(t)}=-\zeta_j(t).             
\eeq
 It follows then from (\ref{E44})-(\ref{E47}) that the master equation,
\beq
(S,S)=0,                                                             
\eeq
holds for $S$, where the notation $( , )$ on the left-hand side  means the so-called
"antibracket" well-known in
the BV formalism \cite{BV} for covariant quantization of gauge field systems,
\beq
(f,g )=\int dtf\left(\frac{\overleftarrow{\delta}}{ \delta z^i(t)}
\frac{\overrightarrow{\delta}}{ \delta z^*_i(t)}-
\frac{\overleftarrow{\delta}}{ \delta z^*_i(t)}
\frac{\overrightarrow{\delta}}{\delta z^i(t)}\right)g.         
\eeq

\section{Discussions}

We have introduced the conception of finite BRST-BFV transformations in the generalized
Hamiltonian formalism for dynamical systems with constraints \cite{FV,BVhf,FF}.
It was shown that the Jacobian of finite BRST-BFV transformations, being the
main ingredient of the approach, can be calculated explicitly in terms of the corresponding
generator acting on finite field-dependent functional parameter of
these transformations. We have introduced the compensation equation providing for a
connection between the generating functionals of Green functions formulated for a given
dynamical system in two different gauges. We have found an explicit solution to the
compensation equation. We have proposed the functional approach to BRST-BFV
transformations, and then reproduced all the results obtained above, in functional terms.
The functional formulation of finite BRST-BFV transformations provided for deriving in
a simple way the Ward identity and connection between the generating functional of Green
functions written in different gauges.

In the present paper we have explored the standard BRST-BFV symmetry in the
generalized Hamiltonian formalism \cite{FV,BVhf,FF}. Some years ago, we have
proposed \cite {BLTh1,BLTh2} the generalized Hamiltonian formalism (which is
also known as the Sp(2)-canonical quantization) based on the extended BRST
symmetry. Instead of one Fermionic parameter in the BRST-BFV transformations,
in the latter case one has to deal with two Fermionic parameters. It seems
very interesting to extend the results obtained above to the case of extended BRST
symmetry.

\section*{Acknowledgments}
\noindent
 I.A.B. would like  to thank Klaus Bering of Masaryk
University for interesting discussions. The work of I.A.B. is
supported in part by the RFBR grants 14-01-00489 and 14-02-01171.
 The work of P.M.L. is
partially supported by the Ministry of Education and Science of
Russian Federation, grant TSPU-122, by the Presidential grant
88.2014.2 for LRSS and  by the RFBR grant 12-02-00121.
The work of I.V.T. is partially supported by the RFBR grant
14-01-00489.
\\

\end{document}